\documentclass[10pt,twocolumn,letterpaper]{article}
\usepackage{cite}
\usepackage{cvpr}
\usepackage{epsfig}
\usepackage{times}  
\usepackage[hyphens]{url}  
\usepackage{graphicx} 
\usepackage{algorithm}
\usepackage{algorithmicx}
\usepackage[noend]{algpseudocode}
\usepackage{color}
\usepackage{amsmath}
\usepackage{amssymb}
\usepackage{amsfonts}
\usepackage{mathrsfs}
\usepackage{mathtools}
\usepackage{bm}
\usepackage{enumitem}
\usepackage{stmaryrd}
\usepackage{multirow}
\usepackage{pifont}
\usepackage[breaklinks=true,bookmarks=false]{hyperref}

\cvprfinalcopy 

\def\cvprPaperID{6669} 

\ifcvprfinal\fi

\DeclareMathOperator{\Enc}{\mathsf{Enc}}
\DeclareMathOperator{\Dec}{\mathsf{Dec}}
\title{ENSEI: Efficient Secure Inference via Frequency-Domain \\
Homomorphic Convolution for Privacy-Preserving Visual Recognition}

\author{
Song Bian$^{1}$\\
\and
Tianchen Wang$^{2}$\\
\and
Masayuki Hiromoto$^{1}$\\
\and
Yiyu Shi$^{2}$\\
\and
Takashi Sato$^{1}$\\
{\tt\small \{sbian, hiromoto, takashi\}@easter.kuee.kyoto-u.ac.jp}\\
{\tt\small \{twang9, yshi4\}@nd.edu}\\
$^{1}$Department of Communications and Computer Engineering, Kyoto University\\
$^{2}$Department of Computer Science and Engineering, University of Notre Dame
}

\begin{document}

\maketitle
\newcommand{\BoldVec}[1]{
  \expandafter\def\csname b#1\endcsname{{\bf{#1}}}%
}
\newcommand{\InnerProd}[2]{
  \expandafter\def\csname in#1#2\endcsname{\text{$\langle{\bf{#1}},{\bf{#2}}\rangle$}}%
}
\newcommand{\mRound}[1]{
  \text{$\left\lfloor #1\right\rceil$}%
}
\newcommand{\numberthis}{
  \addtocounter{equation}{1}\tag{\theequation}
}
\newcommand{\ElementMult}[2]{
  #1\boxcircle #2
}

\newcounter{magicrownumbers}
\newcommand\rownumber{\stepcounter{magicrownumbers}\arabic{magicrownumbers}.}
\BoldVec{x}\BoldVec{y}\BoldVec{a}\BoldVec{s}\BoldVec{c}\BoldVec{r}\BoldVec{p}\BoldVec{e}\BoldVec{b}
\BoldVec{u}\BoldVec{v}\BoldVec{w}\BoldVec{q}\BoldVec{n}\BoldVec{t}\BoldVec{o}
\InnerProd{a}{s}

\newtheorem{claim}{Claim}
\def\zz{{\mathbb{Z}}}
\def\zf{{\mathbb{F}}}
\def\zr{{\mathbb{R}}}
\def\zc{{\mathbb{C}}}
\def\crr{{\mathcal{R}}}
\def\crk{{\mathcal{K}}}
\def\crkgalois{{\mathcal{K_{\rm Galois}}}}
\def\cra{{\mathcal{A}}}
\def\crp{{\mathcal{P}}}
\def\Conv{{\mathsf{Conv}}}
\def\HomConv{{\mathsf{HomConv}}}
\def\FC{{\mathsf{FC}}}
\def\DFT{{\mathsf{DFT}}}
\def\IDFT{{\mathsf{IDFT}}}
\def\NTT{{\mathsf{NTT}}}
\def\INTT{{\mathsf{INTT}}}
\def\NTTD{{\mathsf{NTT2D}}}
\def\INTTD{{\mathsf{INTT2D}}}
\def\FFT{{\mathsf{FFT}}}
\def\IFFT{{\mathsf{IFFT}}}
\def\penc{p_{\rm E}}
\def\pntt{p_{\rm N}}
\def\pfft{p_{\rm F}}
\def\pass{p_{\rm A}}
\def\DFTD{{\mathsf{DFT2D}}}
\def\IDFTD{{\mathsf{IDFT2D}}}
\def\relu{{\mathsf{ReLU}}}
\def\Perm{{\mathsf{Perm}}}
\def\rot{{\mathsf{rot}}}
\def\SIMDScMult{{\mathsf{SIMDScMult}}}
\def\flatten{{\mathsf{flatten}}}
\def\clip{{\mathsf{clip}}}
\def\round{{\mathsf{round}}}
\def\Decomp{{\mathsf{Decomp}}}
\def\share{\mathsf{Share}}
\def\HomShare{\mathsf{HomShare}}
\def\rec{\mathsf{HomRec}}
\def\intFFT{\mathsf{IntFFT2D}}
\def\intIFFT{\mathsf{IntIFFT2D}}

\def\hatu{{\hat{u}}}
\def\hats{{\hat{s}}}
\def\hatw{{\hat{w}}}
\def\ovu{{\overline{u}}}
\def\ovw{{\overline{w}}}
\def\hatU{{\hat{U}}}
\def\hatR{{\hat{R}}}
\def\hatW{{\hat{W}}}
\def\hatbu{{\hat{\bu}}}
\def\hatbr{{\hat{\br}}}
\def\hatbw{{\hat{\bw}}}
\def\hatbv{{\hat{\bv}}}
\def\hatbx{{\hat{\bx}}}
\def\hatby{{\hat{\by}}}
\def\hatbs{{\hat{\bs}}}
\def\bpt{b}
\def\etarot{\eta_{\rm rot}}
\def\etamult{\eta_{\rm mult}}
\def\aaprox{\alpha}
\def\fe{{f_{\rm E}}}
\def\fd{{f_{\rm D}}}
\def\ee{{\varepsilon}}
\def\bee{{\bm{\varepsilon}}}
\def\bpi{{\bm{\pi}}}
\def\etal{et~al.\ }
\def\clgq{\lceil\lg{q}\rceil}
\def\clgp{\lceil\lg{\p}\rceil}
\def\clgpntt{\lceil\lg{\pntt}\rceil}
\def\clgpfft{\lceil\lg{\pfft}\rceil}
\def\clgpenc{\lceil\lg{\penc}\rceil}

\algnewcommand\algorithmicforeach{\textbf{for each}}
\algdef{S}[FOR]{ForEach}[1]{\algorithmicforeach\ #1\ \algorithmicdo}

\begin{abstract}
In this work, we propose ENSEI, a secure inference (SI) framework based on the frequency-domain 
secure convolution (FDSC) protocol for the
efficient execution of privacy-preserving visual recognition.
Our observation is that, under the combination of homomorphic encryption
and secret sharing, homomorphic convolution can be obliviously carried out
in the frequency domain, significantly
simplifying the related computations.  We provide protocol designs and
parameter derivations for number-theoretic transform (NTT) based FDSC.
In the experiment, we thoroughly study the
accuracy-efficiency trade-offs between time- and frequency-domain homomorphic
convolution.  With ENSEI, compared to the best
known works, we achieve 5--11x online time reduction, up to 33x setup time
reduction, and up to 10x reduction in the overall inference time.
A further 33\% of bandwidth reductions can be obtained on binary
neural networks with only 1\% of accuracy degradation on the CIFAR-10
dataset.



\end{abstract}

\sloppy
\section{Introduction}\label{sec:intro}
The design and implementation of privacy-preserving image recognition based on deep
neural network (more generally, secure machine learning as a service (MLaaS)) attract
increasing
attentions~\cite{gilad2016cryptonets,liu2017oblivious,mohassel2017secureml,juvekar2018gazelle,makri2017epic,rouhani2018deepsecure, riazi2019xonn, brutzkus2018low}.
In the field of vision-based MLaaS,  proprietary model 
stealing~\cite{wang2018stealing,juuti2019prada} and privacy
violation~\cite{shokri2017membership} have become one of the main limitations for
the real-world deployment of ML algorithms. For example, serious privacy
concerns have been raised against 
medical imaging~\cite{gomathisankaran2013ensure}, image-based
localization~\cite{speciale2019privacy}, and video
surveillance~\cite{chattopadhyay2007privacycam, wu2018towards}, as 
leaking visual representations in such applications can profoundly undermine the 
well-beings of individuals.
The threat model in a secure inference (SI) scheme for visual recognition can be informally 
formulated as follows. Suppose that Alice as a client wishes to inference
on some of her input images using pre-trained inference engines (e.g., deep
neural networks) from Bob. As the image contains her sensitive information 
(e.g., organ
segmentation), Alice does not want to reveal her inputs to Bob. On the other
hand, it is also financially unwise for Bob who owns the engine to transfer
the pre-trained knowledge base to an untrusted or even malicious client.
A privacy-preserving visual recognition scheme attempts to address the 
privacy and model security risks simultaneously via advanced cryptographic 
constructions, such as homomorphic encryption~\cite{brakerski2012fully, fan2012somewhat} 
and garbled circuits~\cite{yao1982protocols}.

Due to the multidisciplinary nature of the topic, SI
on neural networks (NN) involves contributions from many distinct fields of study.
Initial design explorations mainly focused on the feasibility of NN-based SI, and
generally carry impractical performance
overheads~\cite{gilad2016cryptonets,mohassel2017secureml}. Recent advances in
cryptographic primitives~\cite{halevi2018faster} and adversary
models~\cite{liu2017oblivious,juvekar2018gazelle} have brought input-hiding SI
into practical domain, where $32\times 32$ image datasets can be classified within
seconds~\cite{juvekar2018gazelle, riazi2019xonn}. Meanwhile, from a learning
perspective, alternative feature
representations are discussed to reduce the computational complexity of
SI~\cite{brutzkus2018low}. We also observe that hardware-friendly
network architectures~\cite{courbariaux2015binaryconnect,courbariaux2016binarized} can be
adopted in a secure setting to reduce the computational and communicational
overheads~\cite{riazi2019xonn}.  Furthermore, design optimizations on the
fundamental operations (e.g., secure matrix
multiplication~\cite{jiang2018secure}, secure
convolution~\cite{juvekar2018gazelle}) in SI can also greatly improve its
practical efficiency.

In this work, we propose ENSEI, a general protocol designed for efficient
secure inference over images by adopting homomorphic frequency-domain
convolution (FDC). It is demonstrated that frequency-domain convolution,
which is a key component in reducing the computational complexity of
convolutional neural networks, can also be performed
obliviously, i.e., without the participating parties revealing any piece of
their confidential information.  In addition, we observe that by using the
ENSEI protocol, the complex cryptographic procedure of homomorphic
multiply-accumulate (MAC) operation can also be simplified to efficient
element-wise integer multiplications.  Our main contributions are summarized
as follows.

\begin{itemize}
  \setlength{\parskip}{0cm}
  \setlength{\itemsep}{0cm}
  \item {\bf{Frequency-Domain Secure Inference}}: 
    To the best of our knowledge, we are the first to adopt FDC in
    convolutional neural network (CNN) based
    secure inference. The proposed protocol works for any additive homomorphic
    encryption scheme, including pure garbled circuit (GC) based inference
    schemes~\cite{riazi2019xonn}.
  \item {\bf{NTT-based Homomorphic Convolution with Homomorphic Secret Sharing (HSS)}}:
    The key observation is that FDC can be carried out
    obliviously. Namely, since the discrete Fourier transform operator is
    linear, it can be overlaid with HSS to achieve a weight-hiding
    frequency-domain secure convolution (FDSC). 
    In the experiment, we compare ENSEI-based secure inference with the most recent
    arts~\cite{juvekar2018gazelle, riazi2019xonn}. For convolution benchmarks, we observe 5--11x 
    online and 34x setup time reductions. For whole-network inference time, we
    observe up to 10x reduction, where deeper neural networks enjoy more
    reduction rates.
  \item {\bf{Fine-Grained Architectural Design Trade-Off}}:
    We show that when ENSEI is adopted, different neural architectures with the
    same prediction accuracy can vary significantly in inference time.
    As we observe a 6x performance difference between neural networks with 
    the same prediction accuracy, performance-aware
    architecture design becomes one of the most important
    areas of research for efficient SI.
\end{itemize}

\section{Related Works}\label{sec:related}

{\bf{Secure Inference Based on Interactive Protocols}}:
In MinioNN~\cite{liu2017oblivious}, GC and additive secret sharing (ASS) are
used to transform CNN into oblivious neural networks that ensure data
privacy during secure inference, where a secure inference on one image from
the CIFAR-10 dataset requires more than 500 seconds.
DeepSecure~\cite{rouhani2018deepsecure} further optimize GCs in MinioNN used
in NN layers.  Even with simple dataset such as MNIST, DeepSecure still
requires more than 10 seconds and 791\,MB of network bandwidth.
SecureML~\cite{mohassel2017secureml} adopts the multiplication triples
technique to transfer some computations offline, accelerating the on-line
inference time.  Nevertheless, SecureML requires the existence of two
non-colluding servers, and the inference time is yet from practical.

{\bf{Secure Inference Based on Homomorphic Encryptions}}:
Instead of only using HE to generate multiplication
triples, CryptoNets~\cite{gilad2016cryptonets} and Faster
CryptoNets~\cite{chou2018faster} explores the use of leveled HE (LHE) in
secure inference. With the power of LHE, a two-party protocol is devised
where interactions between the server and the client are minimized.
However, due to the fact that HE parameters scale with the number of network
layers, one of the most recent work~\cite{brutzkus2018low} still requires more
than 700 seconds to evaluate a relatively shallow neural network.

{\bf{The Hybrid Protocol}}: By combining the interactive and homomorphic
approaches, Gazelle~\cite{juvekar2018gazelle} significantly improved the
efficiency of secure inference compared to existing works. The details on
Gazelle are presented later.

\section{Preliminaries}\label{sec:background}

\subsection{Packed Additive Homomorphic Encryption}\label{sec:bfv}
In this work, we focus exclusively on lattice-based PAHE schemes. In
particular, the BFV~\cite{brakerski2012leveled,fan2012somewhat} cryptosystem is
used as it is widely implemented (e.g., SEAL~\cite{chen2017simple},
PALISADE~\cite{palisade}). Here, we give a short overview on the basic
operations of BFV.

THe BFV scheme is parameterized by three variables $(n, \penc, q)$, where $n$
represents the lattice dimension, and is the main security parameter. $\penc$
is the plaintext modulus that determines the maximum size of the plaintext,
and $q$ is the ciphertext modulus.  Similar to
Gazelle~\cite{juvekar2018gazelle}, we use $[\bu]$ to refer to a PAHE
ciphertext holding a plaintext vector $\bu$, where $\bu\in\zz_{\penc}^{n}$. In
lattice-based PAHE schemes, using the Smart-Vercaueren packing
technique~\cite{smart2010fully}, a ciphertext $[\bu]$ can be
constructed from a set of two integer vectors  of dimension $n$ (these vectors
represent the coefficients of some polynomials). Since the ciphertexts are
actually polynomials rather than vectors, for the encryption of a vector
$\bu$, we first turn $\bu$ into a polynomial $u\in\crr_{q}$ where $\crr_{q}$
is some residual ring. The ciphertext in
BFV is then is structured as a vector of two polynomials $[\bu]=(c_0,
c_1)\in\crr^{2}_{q}$, where
\begin{align}\label{eq:bfv_enc}
  c_{0} &= -a, c_{1} = a\cdot t + \frac{q}{\penc}u + e_0. &
\end{align}
Here, $a$ is a uniformly sampled polynomial, and $t, e_0$  are polynomials
whose coefficients are drawn from some discrete Gaussian distributions. 
To decrypt, one simply computes
$\frac{\penc}{q}(c_{0}t+c_1)$ and round off the fractions. 
Note that the multiplications (the ($\cdot$) operation) between polynomials
translate to (nega)cyclic convolutions of the integer vectors representing their
coefficients. For example, in Eq.~\eqref{eq:bfv_enc}, let $\ba,
\bt\in\zz^{n}_{q}$ be
the coefficients of $a$ and $t$, then $a\cdot t=\ba*\bt$, where $*$ is the
convolution operator.

Except for $\Enc$ and $\Dec$ that denote the encryption and decryption
functions, respectively, we define the following three abstract operations for
an AHE scheme. Recall that $[\bx]$ refers the encrypted ciphertext of
$\bx\in\zz^{n}_{\penc}$
\begin{itemize}
  \item Homomorphic addition $(\boxplus)$: for $\bx, \by\in\zz^{n}_{\penc}$, 
    $\Dec([\bx]\boxplus[\by])=\bx+\by$.
  \item Homomorphic Hadamard product $(\boxcircle)$: for $\bx, \by\in\zz^{n}_{\penc}$,
    $\Dec(\ElementMult{[\bx]}{\by})=\bx\circ \by$, where $\circ$ is the
    element-wise multiplication operator.
  \item Homomorphic rotation $(\rot)$: for $\bx\in\zz^{n}_{\penc}$, let
  $\bx=(x_0, x_1, \cdots, x_{n-1})$, 
    $\rot([\bx], k)=(x_{k}, x_{k+1}, \cdots, x_{n-1}, x_{0}, \cdots, x_{k-1})$
  for $k\in\{0, \cdots, n-1\}$.
\end{itemize}
We refer to the efficient implementation of $\boxcircle$ by PAHE schemes as
$\SIMDScMult$. We note that both $\boxplus$ and $\boxcircle$ (when
$\SIMDScMult$ is available) are cheap operations, while $\rot$ and the
homomorphic convolution operation described in Section~\ref{sec:mv_mult} are much more
expensive.


\subsection{Secure Neural Network Inference}\label{sec:gazelle}
\begin{figure}[!t]
 \centering
 \includegraphics[width=0.88\columnwidth]{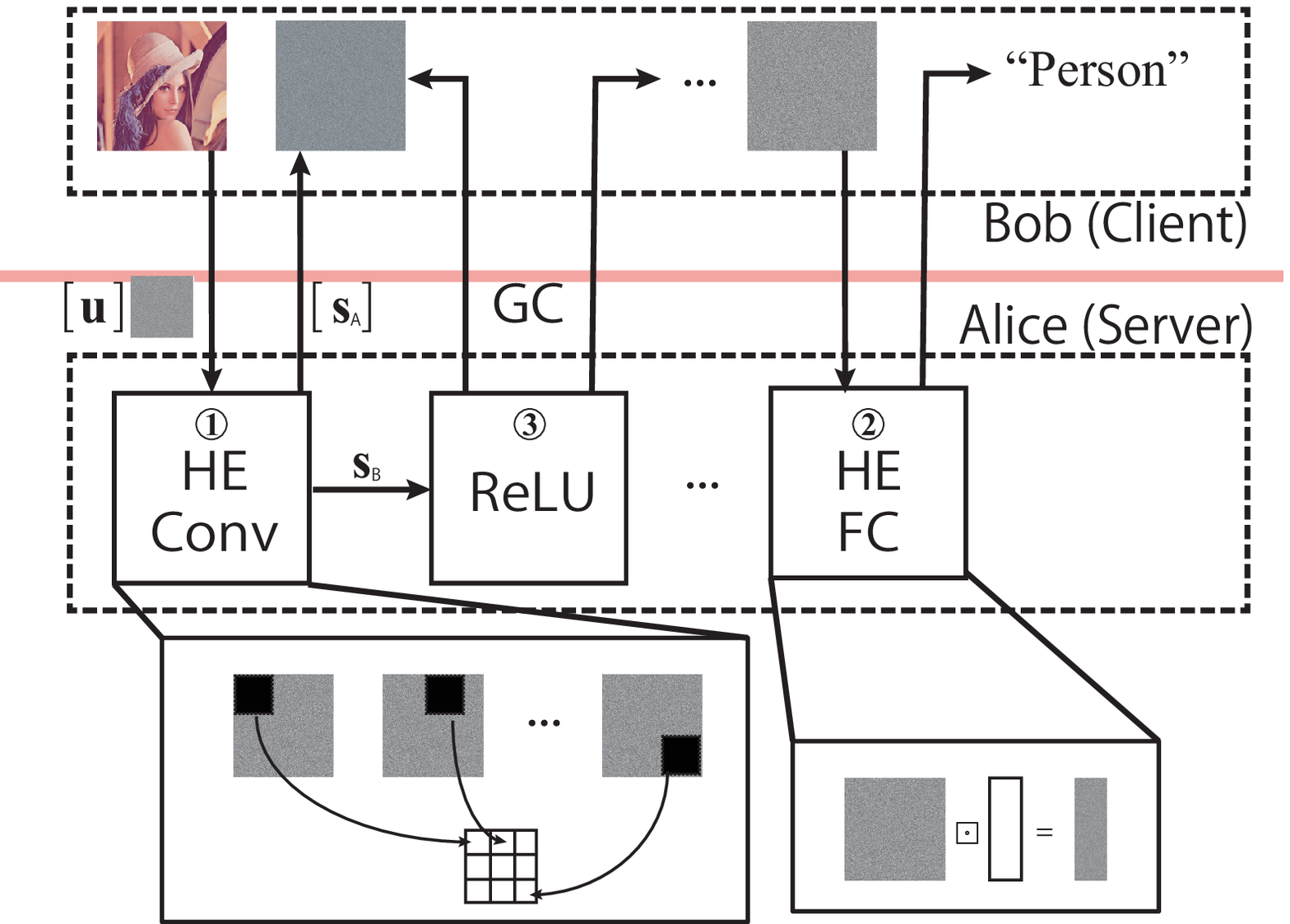}
 \caption{The general architecture of the Gazelle inference engine. The $\FC$
 layer, much like the $\Conv$ layers, is internally a homomorphic
 matrix-vector product.}
 \label{fig:gazelle_gen}
\end{figure}

The main procedures of Gazelle~\cite{juvekar2018gazelle} can be summarized using
the example network architecture shown in Fig.~\ref{fig:gazelle_gen}. As shown
in the figure, the protocol consists of three types of layers: {\large\ding{192}} 
the convolution layer $\Conv$, {\large\ding{193}} the fully-connected layer $\FC$, 
and {\large\ding{194}} the non-linear layers such
as ReLU and image pooling.  Assuming the input image $U$ is of dimension
$n_{o}\times n_{i}$, Alice flattens $U$ as a vector
$\bu_{0}\in\zz_{p}^{n_{u}}$, where $n_u=n_o\cdot n_i$ (i.e., $U$ is raster
scanned into $\bu$). The protocol starts with Alice who encrypts $\bu$
using some PAHE scheme and sends $[\bu]$ to Bob. Upon receiving the
ciphertext, Bob performs the corresponding computations depending on the layer type.\\
{\large\ding{192}} \hspace{1mm} For $\Conv$ layers, Bob homomorphically convolves $[\bu]$
with a plaintext weight filter $W\in\zz^{f_{h}\times f_{w}}_{p}$. The convolved
result, $[\by]=[W*\bu]$, is randomized by a share of secret $\bs_{B}$ as
\begin{align}\label{eq:homshare}
  \HomShare([\by])=([\bs_{A}], \bs_{B}) = ([\by-\bs_{B}\bmod{\pass}], \bs_{B})
\end{align}
with some prime modulus $\pass$. $[\bs_{A}]$ is returned to Alice, and Bob
keeps $\bs_{B}$.\\
{\large\ding{193}} \hspace{1mm} For $\FC$ layers, Bob computes some matrix-vector
product $[\by]=[W\cdot \bu]$ for a plaintext weight matrix $W\in\zz^{n_o\times
n_i}$, and randomizes the result similarly to the $\Conv$ layers. \\
{\large\ding{194}} \hspace{1mm} For non-linear layers, Gazelle evaluates the inputs in three steps:\\
1) obliviously compute $\by=\bs_{A}+\bs_{B}$
to de-randomize $\by$ computed in linear layers using Bob's input $\bs_{B}$
with GC or multiplication triplets,\\
2) compute the non-linear function $f$ (e.g., ReLU or square
from~\cite{liu2017oblivious}) on $\by$, and \\
3) re-randomize the result $f(\by)$ using another share of secret, $\bs_{B, 1}$
from Bob, and output $\bs_{A, 1}=f(\by)-\bs_{B, 1}\bmod{\pass}$ for Alice,
and $\bs_{B, 1}$ for Bob.\\
Upon receiving the outputs from GC, 
Alice encrypts $\bs_{A, 1}$ as $[\bs_{A, 1}]$, and send the ciphertext to Bob.
Bob then computes
\begin{align}\label{eq:homrec}
  [f(\by)]=\rec([\bs_{A}], \bs_{B})=[\bs_{A}+\bs_{B}\bmod{\pass}]
\end{align}
using $[\bs_{A, 1}]$ and $\bs_{B, 1}$, and obtains
$[f(\by)]$. Bob can then start a new round of linear evaluation (steps 
{\large\ding{192}} and {\large\ding{193}}), until all layers are evaluated.

\subsection{Homomorphic Convolution}\label{sec:mv_mult}
One of the main computational bottlenecks in a typical CNN
architecture~\cite{krizhevsky2012imagenet,he2016deep,huang2017densely} is the
evaluations of the large number of $\Conv$ layers. While $\Conv$ only involves
the simple calculation of a series of inner products, the homomorphic version of
inner product is complex to compute. For example, in BFV, for some ciphertext
vector $[\bu]$ and plaintext vector $\bw$, an inner product $[v] =
[\bw\cdot\bu]$ is computed as 
\begin{align}
  [\bv] &= \ElementMult{\bw}{[\bu]}, \text{and }[y] = \sum_{i=1}^{\lg{(n)}}\rot\left([\bv],  \frac{n}{2^i}\right).\label{eq:hconv}
\end{align}
where expensive homomorphic rotations are required to accumulate
the multiplication results.  As it turns out, the $\rot$ operation is expensive in
terms of both computational time and communicational bandwidth. Therefore, one
of the main objectives of ENSEI is to eliminate this complex homomorphic 
rotate-and-accumulate process.

\section{Oblivious Homomorphic Convolution in the Frequency Domain}
\label{sec:propose}

If homomorphic convolution is expensive, a natural question to ask is that if
we can avoid this operation in the first place. The convolution theorem tells
us that for two discrete sequences $\bw$ and $\bu$, there exists a general
transformation $\DFT$ in the form
\begin{align}
  \DFT(\bx)_{k} = \sum_{i=0}^{n_{f}-1}x_i\cdot \omega^{ik},
\end{align}
where the following property holds
\begin{align}\label{eq:f_conv}
  \DFT(\bw*\bu) = \DFT(\bw)\circ \DFT(\bu).
\end{align}
Without loss of generality, we assume that the length of the signals to be
convolved is $n_{f}$, which is basically the filter dimension in secure
inference (i.e., $n_{f}=f_{h}\cdot f_{w}$). Here, $\omega$ is the $n$-th root
of unity in some field $\zf$ (i.e., $\omega^{n}=1$ over $\zf$). As shown
in~\cite{agarwal1975number}, if we choose finite fields as $\zf$, we obtain
the number theoretic transform (NTT), and Eq.~\eqref{eq:f_conv} still holds.
For unencrypted convolution (e.g., frequency-domain convolution
algorithms adopted in CNN libraries), the NTT-based
approach is not particularly attractive in terms of its performance, due to
the additional reductions modulo some large field prime.

We observe a major difference in the encrypted domain. The main benefit for
adopting NTT as the $\DFT$ operator in {\it{secure}} inference is that, in a
cryptographic setting, finite fields are more natural to use than the complex
number field. Most cryptographic primitives that build on established hardness
assumptions live in finite fields, where arithmetic operations do not handle
real numbers (and complex numbers) particularly well.  Therefore, in this
work, we use NTT as our main transformation realization. 



\subsection{ENSEI: The General Protocol}\label{sec:protocol}
\begin{figure*}[!t]
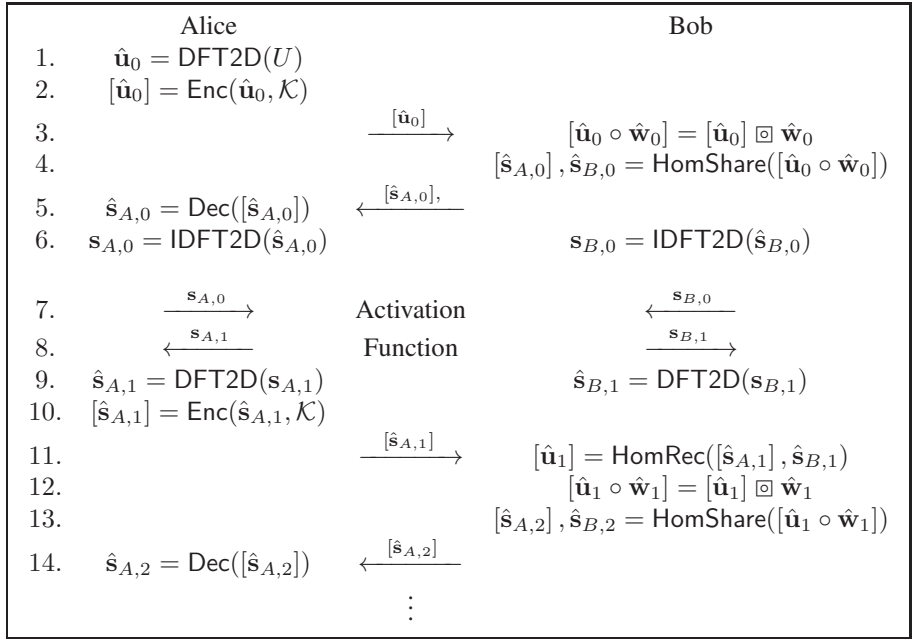

  \centering
  \setcounter{magicrownumbers}{0}
  \[
  \boxed{
    \begin{array}{c c c c}
               & \text{Alice} & & \text{Bob} \\
      \rownumber &      \hatbu_{0} = \DFTD(U) & & \\
      \rownumber &      \left[\hatbu_{0}\right] = \Enc(\hatbu_{0}, \crk) & & \\
    \rownumber & &\xrightarrow{\hspace{0.7em} \left[\hatbu_{0}\right]
    \hspace{0.7em}} & \left[\hatbu_{0}\circ\hatbw_{0}\right] =
      \ElementMult{\left[\hatbu_{0}\right]}{\hatbw_{0}} \\
    \rownumber &                                             & &
      \left[\hatbs_{A, 0}\right], \hatbs_{B, 0} = \HomShare(\left[\hatbu_{0}\circ\hatbw_{0}\right])\\

    \rownumber &      \hatbs_{A, 0} = \Dec(\left[\hatbs_{A, 0}\right]) &\xleftarrow{\hspace{0.7em} \left[\hatbs_{A, 0}\right], \hspace{0.7em}} &  \\
    \rownumber &      \bs_{A, 0} = \IDFTD(\hatbs_{A, 0}) & & \bs_{B, 0} = \IDFTD(\hatbs_{B, 0})\\\\
    \rownumber &      \xrightarrow{\hspace{0.7em} \bs_{A, 0} \hspace{0.7em}} & \text{Activation} & \xleftarrow{\hspace{0.7em} \bs_{B, 0} \hspace{0.7em}}\\
      \rownumber &      \xleftarrow{\hspace{0.7em} \bs_{A, 1} \hspace{0.7em}} & \text{Function} & \xrightarrow{\hspace{0.7em} \bs_{B, 1} \hspace{0.7em}}\\
    \rownumber &      \hatbs_{A, 1} = \DFTD(\bs_{A, 1}) & & \hatbs_{B, 1} =
      \DFTD(\bs_{B, 1}) \\
    \rownumber &      \left[\hatbs_{A, 1}\right] = \Enc(\hatbs_{A, 1}, \crk) & & \\
    \rownumber &                                                                 &\xrightarrow{\hspace{0.7em} \left[\hatbs_{A, 1}\right] \hspace{0.7em}} & \left[\hatbu_{1}\right]=\rec(\left[\hatbs_{A, 1}\right], \hatbs_{B, 1})\\

    \rownumber &                                                                 & & \left[\hatbu_{1}\circ\hatbw_{1}\right] = \ElementMult{\left[\hatbu_{1}\right]}{\hatbw_{1}} \\

    \rownumber &
      & & \left[\hatbs_{A, 2}\right], \hatbs_{B, 2} = \HomShare(\left[\hatbu_{1}\circ\hatbw_{1}\right])\\

    \rownumber &    \hatbs_{A, 2} = \Dec(\left[\hatbs_{A, 2}\right]) &\xleftarrow{\hspace{0.7em} \left[\hatbs_{A, 2}\right] \hspace{0.7em}} &  \\
               & & \vdots &
    \end{array}
  }
  \]
  \caption{The start and intermediate rounds of the Gazelle protocol with
  frequency-domain convolution via $\DFT$. }
  \label{fig:ensei_ntt_protocol}
\end{figure*}

Before delving into the actual protocol, we provide a brief summary of the notations
used in this section to improve the readability of the
derivations.
\begin{itemize}
  \setlength{\parskip}{0cm}
  \setlength{\itemsep}{0cm}
  \item$\bu$, $\bw$: vectors denoting the plaintext input image and weights,
    respectively. When vectors are transformed into the frequency domain, we
    add a hat, e.g., $\hatbu$.

  \item $\bs_{A, 0}$, $\bs_{B, 0}$: vectors referring to the secret shared
  vectors for Alice and Bob, respectively, in the zeroth round of
  communication.

  \item $n_o, n_i, f_h, f_w$: the image and filter dimensions.

  \item  $n, \penc, q$: the RLWE parameters shown in Section~\ref{sec:bfv}.

  \item $\pntt, \pass$: the NTT and the secret sharing moduli, respectively.
\end{itemize}

Figure~\ref{fig:ensei_ntt_protocol} details the general protocol for an
oblivious homomorphic convolution in the frequency domain. In this protocol, we
assume the existence of a pair of general discrete Fourier transform (DFT) operators
$\DFT$ and $\IDFT$, and a pair of homomorphic secret sharing (HSS) scheme
$\HomShare$ and $\rec$ (as depicted in Eq.~\eqref{eq:homshare} and~\eqref{eq:homrec}) 
for randomization and derandomization.
In what follows, we provide a detailed explanation on the proposed protocol.


First, we say that Alice holds some two-dimensional plaintext
input image $U\in\zz^{n_{o}\times n_{i}}$.
Alice wants to inference on $U$ with a set of filters
$\{W\in\zz^{f_{h}\times f_{w}}\}$ held by Bob. The protocol is executed as
follows. 
\begin{enumerate}
  \setlength{\parskip}{0cm}
  \setlength{\itemsep}{0cm}
  \item {\bf{Line~1--3, Alice}}: 
    Alice first pads the input according to the convolution type (e.g., same or
    valid), and computes a two-dimensional DFT on $U$ as $\DFTD(U)$. Here,
    $\DFTD$ is the two-dimensional DFT operator. Alice flattens the frequency-domain
    matrix to ${\hat{\bu}}$ and encrypts it using $\Enc$. The encrypted
    ciphertext $[\hatbu]=\Enc(\hatbu)$ is transferred to Bob through public
    channels.
    

  \item {\bf{Line~3--4, Bob}}: 
    Bob also performs similar padding, DFT, and the flatten operations on the
    (unencrypted) weight matrix, obtaining $\hatbw=\flatten(\DFTD(W))$ as the
    result.  Upon receiving $[\hatbu]$, Bob carries out the frequency-domain
    homomorphic convolution by computing a simple homomorphic Hadamard product
    between $[\hatbu]$ and transformed plaintext $\hatbw$.
    In order to prevent weight leakages, Bob applies $\HomShare$ from the HSS
    protocol and acquires two shares of secrets $[\hatbs_{A, 0}]$ and
    $\hatbs_{B, 0}$ according to Eq.~\eqref{eq:homshare}.
    Bob sends the shared secret $[\hatbs_{A, 0}]$ back to Alice.

  \item {\bf{Line~4--6, Alice}}: 
    Alice decrypts $[\hatbs_{A, 0}]$ as $\hatbs_{A, 0}$. Both Alice and Bob
    apply $\IDFT$ to the shares of secret to obtain
    \begin{align}
      \bs_{A, 0}&=\IDFT(\hatbs_{A, 0})&\nonumber\\
                &=\IDFT(\hatbu_{0}\circ\hatbw_{0}-\hatbs_{B, 0})\bmod{\pass}&\label{eq:idft_result}\\
      \bs_{B, 0}&=\IDFT(\hatbs_{B, 0})\bmod{\pass},& 
    \end{align}
    marking the end of the evaluation for the first convolution layer
    in the NN. We note that when the $\IDFT$ operator is error free,
    Eq.~\eqref{eq:idft_result} evaluates to 
    \begin{align}
      &\IDFT(\hatbu_{0}\circ\hatbw_{0}-\hatbs_{B, 0})\bmod{\pass}&\nonumber\\
      &= \IDFT(\hatbu_{0}\circ\hatbw_{0})-\IDFT(\hatbs_{B, 0})\bmod{\pass}&\label{eq:idft_convolution0}\\
      &= \bu_{0}*\bw_{0}-\bs_{B, 0}\bmod{\pass}\label{eq:idft_convolution1}.
    \end{align}

  \item {\bf{Line~7--8, Alice and Bob}}: 
    In implementing oblivoius activation, the essential computations involved in GC
    (or multiplication triples) can be formulated as
    \begin{align}
      \bu_{0}*\bw_{0} &= \bs_{A, 0}+\bs_{B, 0}\bmod{\pass}&\\
      \bu_{1} &= f(\bu_{0}*\bw_{0})&\\
      \bs_{A, 1} &= \bu_{1}-\bs_{B, 1}\bmod{\pass},&
    \end{align}
    where $f$ is some activation function (e.g., ReLU).
    As mentioned, the above procedure only remains correct if the underlying
    $\DFT$ operator satisfies Eq.~\eqref{eq:idft_convolution0}.

    \item {\bf{Line~9--10, Alice}}:
      Upon receiving the computed results from oblivious activation, 
      Alice repeats the process
      of Line~1--3. The transformed and encrypted inputs
      $[\hatbs_{A, 1}]$ are again sent to Bob.
    \item {\bf{Line~11--13, Bob}}:
      The final step to complete the proposed protocol is the  $\rec$
      procedure as in Eq.~\eqref{eq:homrec}. 
      Since $[\hatbs_{A, 1}]=[\DFT(\bu_{1}-\bs_{B, 1})]$ and
      $\hatbs_{B, 1}=\DFT(\bs_{B, 1})$, when the same condition in
      Eq.~\eqref{eq:idft_convolution0} holds, we have
      \begin{align}
        &[\DFT(\bu_{1}-\bs_{B, 1})]\boxplus \DFT(\bs_{B, 1})&\label{eq:dft_convolution0}\\
        &=[\hatbu_{1}-\DFT(\bs_{B, 1}) + \DFT(\bs_{B, 1})]=[\hatbu_{1}]\label{eq:dft_convolution1}&
      \end{align}
      and a new round of FDSC can be carried out.
\end{enumerate}
The evaluation of any CNN can be decomposed into the repetitions of the above
procedures, as our protocol is essentially a way of obliviously moving into
and out of the frequency domain.


Note that the DFT operator in Fig.~\ref{fig:ensei_ntt_protocol} does not have
to be NTT. However, adopting NTT in the protocol requires minimal modification,
as we only need to replace the $\DFT$ and $\IDFT$ operators
in Fig.~\ref{fig:ensei_ntt_protocol} with $\NTT$ and $\INTT$. In this case,
in addition to the secret sharing modulus $\pass$ and encryption modulus
$\penc$, we need a third modulus $\pntt$ for ENSEI-NTT. In other words, we have
$\omega^{n}\equiv 1\bmod{\pntt}$, and the corresponding transformation is
written as
\begin{align}\label{eq:ntt_transform_eq}
  \NTT(\bx)_{k}=\sum_{i=0}^{n_{f}-1}x_i\omega^{ik}\bmod{\pntt}.
\end{align}

\subsubsection{Correctness for NTT-based ENSEI}
Here, the correctness of the NTT version of our protocol is briefly explained, 
and a more detailed
discussion can be found in the appendix.  We assert that due to the need of
finite-field arithmetic, not every pair of $\DFT$ operators work for the above
protocol.  The most important condition to ensure correctness is that
Eq.~\eqref{eq:idft_convolution0} equals to Eq.~\eqref{eq:idft_convolution1},
and that Eq.~\eqref{eq:dft_convolution0} equals to
Eq.~\eqref{eq:dft_convolution1}. If we instantiate ENSEI with NTT, then, the
correctness holds when the following statement is true
\begin{align}
&\INTT(\hatbu\circ\hatbw-\hatbs_{B})\bmod{\pass}&\\
  &=(\INTT(\hatbu\circ\hatbw)-\INTT(\hatbs_{B}))\bmod{\pass}&\\
  &=(\bu*\bw - \bs_{B})\bmod{\pass}.&
\end{align}
The convolution result $\bu*\bw$ can be recovered by applying the recovery
procedure of HSS, under the condition that the HSS modulus $\pass$ is larger
than the NTT modulus $\pntt$. A similar procedure also ensures the correctness
of Eq.~\eqref{eq:dft_convolution0} and Eq.~\eqref{eq:dft_convolution1}. 


\subsubsection{Security}
In terms of security properties, our protocol is basically identical to the
linear kernel in Gazelle~\cite{juvekar2018gazelle}, so a formal proof is left
out. Briefly speaking, given the security of the PAHE scheme, Bob cannot temper
the encrypted inputs of Alice (e.g., $\bu$), and with HSS, Alice gains no knowledge of
the models from Bob (e.g., $\bw$).

\section{Integration and Parameter Instantiation}\label{sec:integr}


\subsection{Reducing the Number of DFT in ENSEI}\label{sec:nontt}
\begin{figure}[!t]
 \centering
 \includegraphics[width=1\columnwidth]{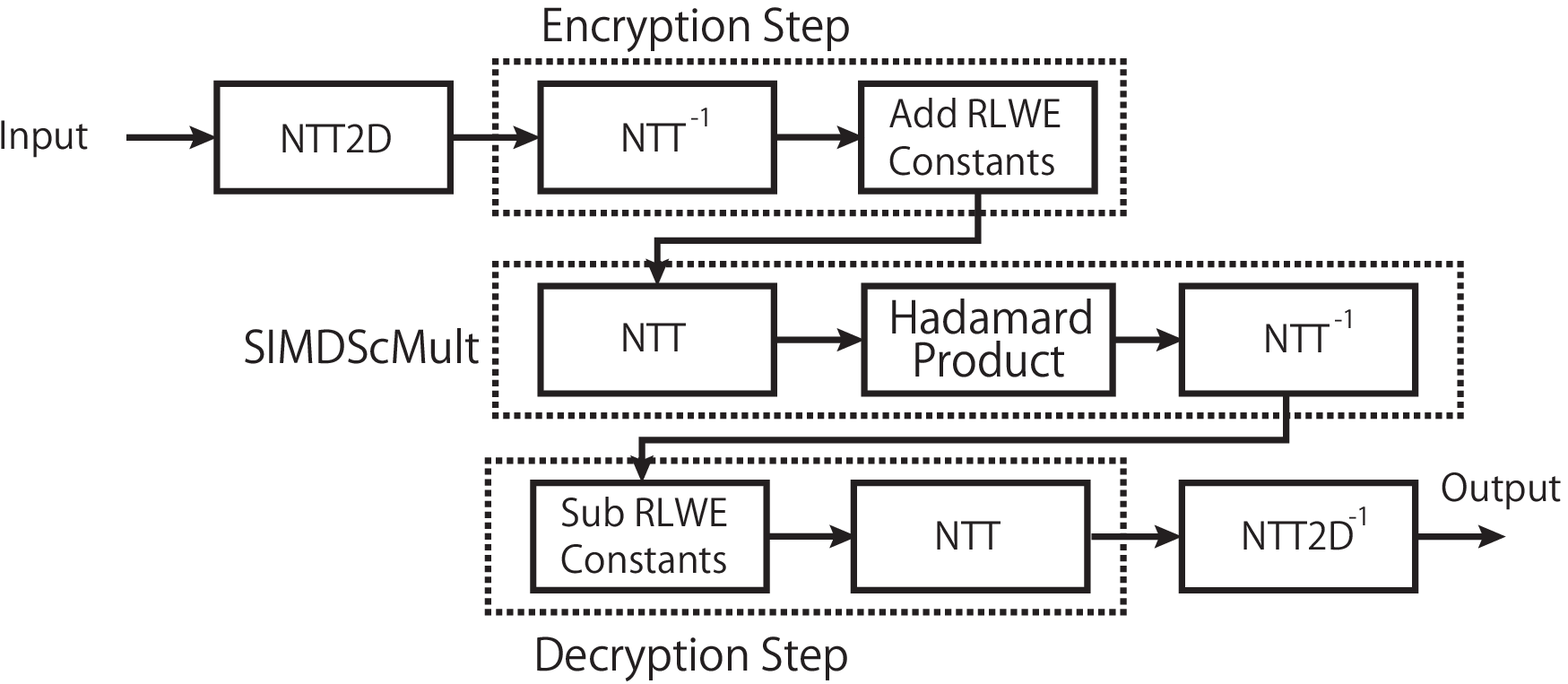}
 \caption{The overview for a sequence of $\Enc$-$\SIMDScMult$-$\Dec$
 procedures based on ENSEI for general AHE schemes.}
 \label{fig:nontt}
\end{figure}

\begin{figure}[!t]
 \centering
 \includegraphics[width=1\columnwidth]{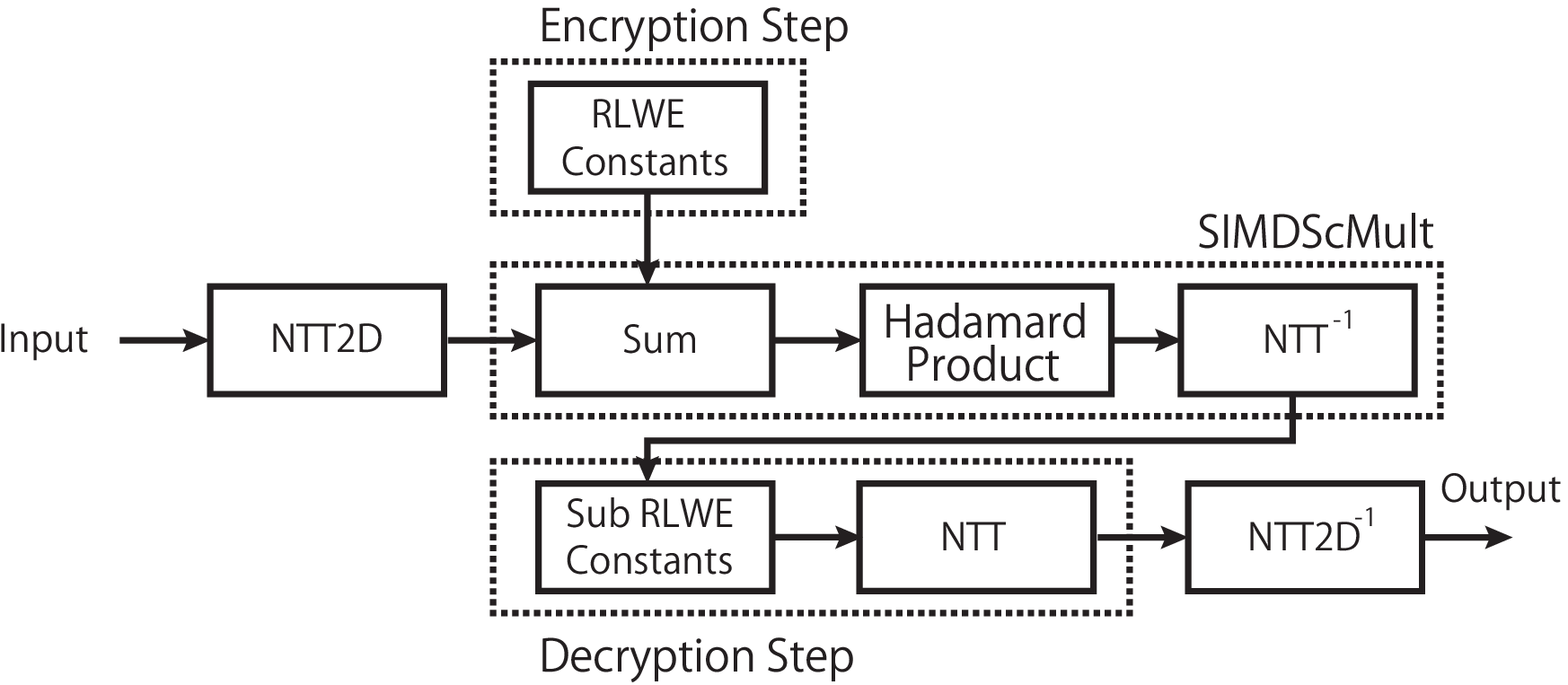}
 \caption{Modified $\Enc$-$\SIMDScMult$-$\Dec$ for Gazelle-like networks to
 reduce the extra NTTs for plaintext packing.}
 \label{fig:nontt_r}
\end{figure}
The plaintext packing technique~\cite{smart2010fully} used for embedding a
vector of plaintext integers $\bu\in\zz^{n}_{p}$ into a single ciphertext pair
relies on the idea that a large-degree polynomials (with proper modulus) can
be decomposed into a set of independent polynomials that are of smaller
degrees. The exact procedures for lattice-based PAHE is sketched in
Fig.~\ref{fig:nontt}, where we can see that during encryption, a vector of
plaintext is transformed into the time domain via the $\INTT$ operation and
embedded into the ciphertext. Later in the evaluation stage, $\SIMDScMult$
re-applies $\NTT$ on the ciphertext (and thus simultaneously on the plaintext)
followed by an element-wise multiplication, which also conducts the
coefficient-wise multiplication on the plaintext vector. We see that while
Fig.~\ref{fig:nontt} is a straightforward application of ENSEI, it clearly involves redundant
NTTs.

Our key observation here is that, the internal operation of a $\SIMDScMult$ is
merely conducting frequency-domain multiplication of polynomials (shown in
Fig.~\ref{fig:nontt}), as it is known that a multiplication between two
polynomials in some particular quotient rings equates to a (nega-)cyclic
convolution of their coefficients.  Therefore, we can directly embed the
NTT2D-transformed frequency-domain image into the NTT transformed ciphertext
(i.e., $\hat{a}\cdot\hat{s}+\NTTD(U)$), as depicted in Fig.~\ref{fig:nontt_r},
and execute $\SIMDScMult$ without NTT operations during the encryption stage
(the RLWE constants can be generated offline).
By performing the entire $\Enc$-$\SIMDScMult$ process in the
frequency domain, we can reduce two NTT butterflies per
convolution. Note that, because the homomorphic rotations employed in Gazelle 
($\rot$) force several rounds of NTTs for switching the
decryption keys (such that proper keys are generated to decrypt the rotated 
ciphertext), ENSEI needs much less NTT runs compared to the 
time-domain convolution devised by Gazelle. The only restriction in ENSEI is 
that the plaintext modulus 
$\penc$ needs to be larger than the ENSEI modulus $\pntt$.  
Further elaborations on the exact precision settings that
satisfy the requirement $\penc\geq \pntt$ is discussed in the experiment. 



\subsection{The Moduli and RLWE Parameters}\label{sec:precision}

In ENSEI, we have three moduli to consider: the secret sharing modulus
$\pass$, the encryption modulus $\penc$, and the NTT modulus $\pntt$. The
three moduli need to satisfy the relation $\penc\geq \pass\geq \pntt$. As it turns out,
$\pntt$ is determined by two factors: i) the maximum value in
the matrix operands, and ii) the length of the convolving sequence.
For two sequences $\bu\in\zz^{n_{o}\times n_{i}}$ and $\bw\in\zz^{f_{h}\cdot f_{w}}$, the lower
bound on $\pntt$ can be written as
\begin{align}\label{eq:eq_pntt}
  \pntt\geq \max(\bu)\cdot\max(\bw)\cdot f_{h}\cdot f_{w}.
\end{align}
Suppose $n_{f}=f_{h}\cdot f_{w}$, 
in a typical CNN setting, compared to the
RLWE lattice dimension $n$, $n_{f}$ is generally small (e.g., $n_{f}=9$ for the
$3\times 3$ filters used in the experiment).  
In addition, in
hardware-friendly network architectures, such as
BinaryConnect~\cite{courbariaux2015binaryconnect} or
BinaryNet~\cite{courbariaux2016binarized}, $\max(\bu)$ is generally less than
10-bit, and $\max(\bw)$ is even smaller. 

As described, $\penc$ only needs to be as large as $\pntt$. This is not a
problem when $\pntt$ is large. However, when all the terms in
Eq.~\eqref{eq:eq_pntt} is small, we can set $\pntt$ to be extremely small, but
not $\penc$.  For security reasons, $n$ needs to be a relative large power of
2 (e.g., 1024 to 2048), and $\penc$ can only be as small as the smallest prime
that completely splits over the field $x^{n}+1$ (e.g., for $n=2048$,
$\penc\geq 12289$), and we cannot set $\penc=\pntt$ if $\pntt<12289$.
Fortunately in this case, since $\pntt$ is not related to the security of
PAHE, we can still use a smaller $\pntt$ to transform the weight matrix. The
small size of $\pntt$ makes the coefficients of the transformed weight matrix
small, thereby reducing the noise growth and the size of the ciphertext
modulus. 

\section{Complexity Analysis and Numerical Experiments}\label{sec:experiment}
\subsection{Complexity Analysis for ENSEI}
Given the integrated protocol, we give a comparison between the asymptotic
computational complexity of Gazelle and ENSEI. We formulate the
complexity based on three basic operations, $t_{\penc}$, $t_{q}$, and
$t_{\pntt}$, which are the respective time of performing a 
multiplication modulo $\penc$, $q$, and $\pntt$. As described above, we assume the
convolution is performed between the input $U$ of dimension $n_{u}=n_{o}\times
n_{i}$, and $W$ of dimension $n_{f}=f_{h}\times f_{w}$. We also use
$\beta_{n_{u}}$ for $\lceil \frac{n_{u}}{n}\rceil$, the ratio between the
input image dimension and lattice dimension. For ENSEI-NTT, we have
\begin{align}\label{eq:complexity_ntt}
  \beta_{n_{u}}\big(4\cdot (n\cdot \log(n_{u})\cdot t_{\pntt})+ n \cdot t_{q}
  + 2\cdot n\log(n)\cdot t_{q}\big),
\end{align}
which is of order $\mathcal{O}(n_{u}\cdot \log(n_{u}))$, since the RLWE parameter $n$ is a
constant.  In Eq.~\eqref{eq:complexity_ntt}, the
forward NTTs on $U$ and $W$, and the backward inverse NTT on the convolution
result and randomization vector counts for four transformations over the field
$\zf_{\pntt}$. The last term in Eq.~\eqref{eq:complexity_ntt} counts for the
encryption and decryption costs, where NTTs are performed on the ciphertext
with lattice dimension $n$.  Since Gazelle~\cite{juvekar2018gazelle} did not
provide formal complexity calculations, the analyses here are only of
estimations.
\begin{align}\label{eq:complexity_gazelle}
  &\beta_{n_{u}}\big(2\cdot n\log(n)\cdot t_{q}+ (n_{i}\cdot f_{w}-1)\cdot (n_{o}\cdot f_{h}-1)\cdot n\cdot t_{q} +&\nonumber\\
  &n\log(n)\cdot t_{q} + n\log(n)\cdot t_{Q}\big),&
\end{align}
where $t_{Q}$ is the time for generating the Galois key with a slightly larger
modulus $Q$, as described in~\cite{halevi2018faster}. Two points are
emphasized here. First, it can be seen that the second term in
Eq.~\eqref{eq:complexity_gazelle} depends on both the input image and filter
window, which means that the complexity is of
$\mathcal{O}(n_{u}\cdot\log(n_{u})\cdot n_{f})$.  In contrast, ENSEI-NTT is
only of $\mathcal{O}(n_{u}\cdot\log(n_{u}))$.  
This is the primary reason why
Gazelle does not scale well, as both the time for rotation
and the ciphertext modulus $q=\mathcal{O}(\mathsf{Poly}(\penc))$ increase as a result
of larger weight matrices.
\begin{table}[!t]
  \footnotesize
  \centering
  \caption{Prediction Accuracy on the CIFAR-10 Dataset}
  \label{tab:accuracy_result}
  \begin{tabular}{c|c|c|c}
  \hline
  Precision  & Input        & Filter       & Accuracy  \\
             & Bit Width    & Bit Width    &           \\\hline
  Binary     & 8            & 1            & 81\%      \\\hline
  Medium     & 8            & 4            & 79\%      \\\hline
  High       & 11           & 7            & 82\%      \\\hline
  High-Square& 11           & 7            & 82\%      \\\hline
  Full       & 32           & 32           & 83\%      \\\hline\hline
  \cite{liu2017oblivious, juvekar2018gazelle} & - & -          & 82\%      \\\hline
  \end{tabular}
\end{table}

\begin{table}[!t]
  \footnotesize
  \centering
  \caption{Proposed Candidate Parameter Sets}
  \label{tab:ensei_params}
  \begin{tabular}{c|c|c|c|c}
  \hline
  Protocol                   & Parameter & Binary          & Medium           & High     \\\hline
  \multirow{6}{*}{ENSEI-NTT} & $\pntt$    & 2311             & 147457            & 2359303  \\
                             & $\clgpntt$ & 11              & 12                & 18       \\
                             & $\penc$    & 12289          & 147457            & 2363393  \\
                             & $\clgpenc$ & 14             & 18                & 22       \\
                             & $\clgq$    & 45             & 53                & 60      \\
   & $n$        & 2048 & 2048 & 2048 \\\hline
  \end{tabular}
\end{table}
\begin{table*}[!t]
  \footnotesize
  \centering
  \caption{Computational Cost of ENSEI with Respect to Different Parameter Sets Measured in Microseconds}
  \label{tab:ensei_performance}
  \begin{tabular}{c|c|c|c|c|c|c|c|c|c|c}
  \hline
                   & Precision& Image & Encryption       & Filtering      & Result     & $t_{\rm online}$  & PAHE      & Filter     & $t_{\rm setup}$ & Galois-Key \\
                   &          & Read  & Constants        & (Hadamard) & Decryption &                  & Setup     &    Read    &                  & Generation \\\hline 
  ENSEI-NTT        & Binary   & 197.4            & 92.1       & 146.7 (7.3)   & 150.6 & 587.1  & 21972.2  & 1l0.6 & 22082.8  & 76983.2 \\\hline
  ENSEI-NTT        & High     & 197.3            & 92.4       & 147.5 (7.3)   & 151.2 & 588.5  & 21351.6  & 110.2 & 21462.2  & 77039.3 \\\hline
 \end{tabular}
\end{table*}

\begin{table*}[!t]
  \footnotesize
  \centering
  \caption{Convolution Benchmarks w.r.t Precision Levels}
  \label{tab:ensei_comparison}
  \begin{tabular}{c|c|c|c|c|c|c}
  \hline
  & Input Dim.               & Filter Dim.            & Precision & $t_{\rm
setup}$ & $t_{\rm online}$ & Bandwidth\\\hline 
  Gazelle & $(28\times 28\times 1)$  & $(5\times 5\times 5)$  & -         & 11.4\,ms          & 9.20\,ms & 130\,KB     \\\hline
  Gazelle & $(32\times 32\times 32)$ & $(3\times 3\times 32)$ & -         &
  704\,ms           & 195\,ms  & -           \\\hline\hline
  ENSEI-NTT & $(28\times 28\times 1)$  & $(5\times 5\times 5)$  & Binary    & 22.0\,ms          & 1.8\,ms & 23.0\,KB    \\\hline
  ENSEI-NTT & $(28\times 28\times 1)$  & $(5\times 5\times 5)$  & High      & 21.4\,ms          & 1.8\,ms & 30.1\,KB    \\\hline
  ENSEI-NTT & $(32\times 32\times 32)$ & $(3\times 3\times 32)$ & Binary    & 22.0\,ms          & 16.9\,ms & 368\,KB     \\\hline
  ENSEI-NTT & $(32\times 32\times 32)$ & $(3\times 3\times 32)$ & High      & 21.4\,ms          & 16.9\,ms & 492\,KB     \\\hline
 \end{tabular}
\end{table*}

\subsection{Experiment Setup}
In order to quantitatively assess the impact of ENSEI on secure inference, we
implemented the ENSEI protocol using the SEAL library~\cite{chen2017simple} in
C++. We also performed accuracy test to estimate the smallest NTT modulus
$\pntt$ for different parametrizations of PAHE.

Most existing works~\cite{juvekar2018gazelle, liu2017oblivious}
only show the main accuracy and performance results on either MNIST or CIFAR-10,
or even smaller datasets~\cite{riazi2019xonn}. For fair comparisons,
we first report accuracy and quantization experiments using the same architecture
in one of the most recent works~\cite{juvekar2018gazelle}, and then compare 
ENSEI-based SI on other architectures (with different protocols) suggested 
in~\cite{riazi2019xonn}.
The accuracy results are obtained using the Tensorflow
library~\cite{tensorflow2015-whitepaper}, and the runtime of homomorphic
convolution is recorded on an Intel Core i3-7100 CPU 3.90\,GHz processor.


\subsection{CNN Prediction Accuracy}\label{sec:accuracy}

Using the same architecture in~\cite{juvekar2018gazelle, liu2017oblivious}, 
Table~\ref{tab:accuracy_result} illustrates how the prediction accuracy 
improves as the bit precision increases in SI. We observe that for 11-bit features
and 7-bit weights, the accuracy is on an equivalent level to the
full-bit precision case (the difference is less than 1\%). Meanwhile, 
the binary-weight instance can reach
a final prediction accuracy of 81\%, which is only 1\% less than the
original accuracy reported in~\cite{juvekar2018gazelle, liu2017oblivious}
(we may have used different hyperparameters).

As later observed in Section~\ref{sec:arch_comp}, using GC-based ReLU on the 
first several convolution layers brings significant performance overhead to
both existing works~\cite{juvekar2018gazelle, liu2017oblivious} and ENSEI. 
Therefore, we experimented on a different network architecture, where some 
of the ReLU layers are replaced with the square activation (SA) function proposed 
in~\cite{liu2017oblivious}. We found that replacing a small amount of ReLU with SA
(denoted as High-Square in Table~\ref{tab:accuracy_result})
does not affect the prediction accuracy much, while 
replacing all ReLU layers with SA does (accuracy becomes only 10\%).
Nonetheless, even replacing a small portion of ReLU with SA proves to be critical
in improving the practicality of ENSEI-based SI, as demonstrated in 
Section~\ref{sec:arch_comp}.



\subsection{Quantization and Parameter Instantiations}\label{sec:quantizatino}

From the previous section, we obtain the maximum values on $\bu$ and
$\bw$. Hence, we can apply Eq.~\eqref{eq:eq_pntt} to instantiate three sets of
RLWE parameters for adopting the fine-grained layer-to-layer (and
network-to-network) precision adjustment. We take the high-precision case with
a $3\times 3$ filter size as an example. For 12-bit $\bu$ and 6-bit $\bw$,
$\pntt$ needs to satisfy
\begin{align}
  \pntt\geq 2^{12}\cdot 2^{6}\cdot 9,
\end{align}
and that $\pntt\equiv 1\bmod{3}$. The lattice dimension $n$ is fixed to be
2048 to ensure efficient packing and a 128-bit security. We find the minimal
$\penc\geq \pntt$ for the $\pntt$ required such that $\penc\equiv 1\bmod{n}$.
Finally, the ciphertext modulus $q$ is adjusted accordingly to
tolerate the error growth while retaining the security requirement, with
overwhelming decryption success probability (the discrete Gaussian parameter
$\sigma$ is set to 4).

The calculated moduli and instantiated parameters are shown in
Table~\ref{tab:ensei_params}. 
Using a $\pntt$ that is much smaller than $\penc$, we generate less
noises in the ciphertext, and the Binary parameter set enjoys from a 
smaller ciphertext modulus.
The resulting ciphertext is 33\% smaller than the High parameter set, as shown 
in Section~\ref{sec:gazelle_comp}.

\subsection{Efficiency Comparison to Gazelle}\label{sec:gazelle_comp}

We summarize the performance data of ENSEI-NTT with respect to
different parameter instantiations in Table~\ref{tab:ensei_performance}. The
running times are an average of 10,000 trials measured in microseconds. Here,
$t_{\rm setup}$ is time consumed by procedures that do not
involve user inputs. Likewise, $t_{\rm online}$ refers to the time for
input-dependent steps, and is the sum of all terms in
Table~\ref{tab:ensei_performance} up to $t_{\rm online}$ horizontally.
In particular, the results show that the time it takes to compute the 
Hadamard product (shown in parenthesis after the Filter column) is only
a fraction of the time consumed by the NTT operations (the filtering step
contains two NTT butterflies), which need to be applied once per each of the input 
and filter channels.

Using the instantiated parameters and recorded speed, the performance
comparison of ENSEI-NTT and Gazelle on a set of convolution benchmarks are
summarized in Table~\ref{tab:ensei_comparison}. In
Table~\ref{tab:ensei_comparison}, we see that for larger benchmarks,
the online convolution time
is reduced by nearly 11x across all precisions with ENSEI-NTT.  
In addition, we point out that the setup time of ENSEI scales
extremely slowly with the dimensions of the images and filters. Therefore,
we observe a nearly 34x reduction in setup time for the larger benchmarks.
In combined, ENSEI-NTT obtains a 23x reduction in total time for a 32-channel
convolution.  

\subsection{Architectural Comparisons}\label{sec:arch_comp}
\begin{table}[!t]
  \footnotesize
  \centering
  \caption{The Impact of Neural Architecture on Inference Time}
  \label{tab:arch_comp}
  \begin{tabular}{c|c|c|c|c}
  \hline
  Architecture                        & \#Conv          & Accuracy  & ENSEI & Prior Arts\\
  & &  & Total Time & Time \\\hline
  Fig.~13 in \cite{liu2017oblivious}  & 7               & 82\%      & 7.72\,s  &  12.9\,s  \\
  High-Square                         & 7               & 82\%      & 1.38\,s  &  -        \\\hline
  BC2 in \cite{riazi2019xonn}         & 9               & 82\%      & 2.76\,s  &  4.8\,s   \\\hline
  BC3 in \cite{riazi2019xonn}         & 9               & 86\%      & 14.7\,s  &  35.8\,s  \\\hline
  BC4 in \cite{riazi2019xonn}         & 11              & 88\%      & 30.91\,s &  123.9\,s \\\hline
  BC5 in \cite{riazi2019xonn}         & 17              & 88\%      & 30.98\,s  &  147.7\,s \\\hline
  \end{tabular}
\end{table}
Lastly, we compare different architectures with and without FDSC to
demonstrate the effectiveness of the ENSEI protocol.
Table~\ref{tab:arch_comp} summarizes the inference time with respect to
increasingly deep neural architectures, and the accuracy is measured on the
CIFAR-10 dataset. We first observe that, while HE-based linear layers
represent a large portion of the computational time (from 40\% up to 80\% of
the total inference time), the GC-based non-linear ReLU computations become
the bottleneck when ENSEI is employed.  By replacing the second and fifth ReLU
layer in the benchmark architecture (the complete architecture can be found in
the appendix), the online SI time can be reduced to 1.38 seconds, nearly 10x
faster than the baseline method. 
Using ENSEI, what we discovered is that, under the same
accuracy constraint,  certain neural architectures are much more efficient
when implemented using secure protocols. Hence, efficient ways of finding such
architecture becomes an important future area of research.

\section{Conclusion}\label{sec:conclusion}
In this work, we proposed ENSEI, a frequency-domain convolution technique that
accelerates CNN-based secure inference. By using a generic DFT, we show that
oblivious convolution can be built on any encryption scheme that is additively
homomorphic. In particular, we instantiate and integrate ENSEI with NTT and
compare ENSEI-NTT to one of the most recent work on secure inference, Gazelle.
In the experiment, we observed up to 23x reduction in convolution time,
and up to 10x in the overall inference time. We demonstrate that PAHE-based
protocol is one of the simplest and most practical secure inference scheme.

\section*{Acknowledgment}
The authors thank Rongyanqi Wang for the discussions and supports.  This work
was partially supported by JSPS KAKENHI Grant No.~17H01713, 17J06952,
Grant-in-aid for JSPS Fellow (DC1), National Science Foundation under Grant
CNS-1822099, and Edgecortix Inc.

{\small
\bibliographystyle{ieee_fullname}
\bibliography{cad,security}
}

\end{document}


\maketitle
\newcommand{\BoldVec}[1]{
  \expandafter\def\csname b#1\endcsname{{\bf{#1}}}%
}
\newcommand{\InnerProd}[2]{
  \expandafter\def\csname in#1#2\endcsname{\text{$\langle{\bf{#1}},{\bf{#2}}\rangle$}}%
}
\newcommand{\mRound}[1]{
  \text{$\left\lfloor #1\right\rceil$}%
}
\newcommand{\numberthis}{
  \addtocounter{equation}{1}\tag{\theequation}
}
\newcommand{\ElementMult}[2]{
  #1\boxcircle #2
}

\newcounter{magicrownumbers}
\newcommand\rownumber{\stepcounter{magicrownumbers}\arabic{magicrownumbers}.}
\BoldVec{x}\BoldVec{y}\BoldVec{a}\BoldVec{s}\BoldVec{c}\BoldVec{r}\BoldVec{p}\BoldVec{e}\BoldVec{b}
\BoldVec{u}\BoldVec{v}\BoldVec{w}\BoldVec{q}\BoldVec{n}\BoldVec{t}\BoldVec{o}
\InnerProd{a}{s}

\newtheorem{claim}{Claim}
\def\zz{{\mathbb{Z}}}
\def\zf{{\mathbb{F}}}
\def\zr{{\mathbb{R}}}
\def\zc{{\mathbb{C}}}
\def\crr{{\mathcal{R}}}
\def\crk{{\mathcal{K}}}
\def\crkgalois{{\mathcal{K_{\rm Galois}}}}
\def\cra{{\mathcal{A}}}
\def\crp{{\mathcal{P}}}
\def\Conv{{\mathsf{Conv}}}
\def\HomConv{{\mathsf{HomConv}}}
\def\FC{{\mathsf{FC}}}
\def\DFT{{\mathsf{DFT}}}
\def\IDFT{{\mathsf{IDFT}}}
\def\NTT{{\mathsf{NTT}}}
\def\INTT{{\mathsf{INTT}}}
\def\NTTD{{\mathsf{NTT2D}}}
\def\INTTD{{\mathsf{INTT2D}}}
\def\FFT{{\mathsf{FFT}}}
\def\IFFT{{\mathsf{IFFT}}}
\def\penc{p_{\rm E}}
\def\pntt{p_{\rm N}}
\def\pfft{p_{\rm F}}
\def\pass{p_{\rm A}}
\def\DFTD{{\mathsf{DFT2D}}}
\def\IDFTD{{\mathsf{IDFT2D}}}
\def\relu{{\mathsf{ReLU}}}
\def\Perm{{\mathsf{Perm}}}
\def\rot{{\mathsf{rot}}}
\def\SIMDScMult{{\mathsf{SIMDScMult}}}
\def\flatten{{\mathsf{flatten}}}
\def\clip{{\mathsf{clip}}}
\def\round{{\mathsf{round}}}
\def\Decomp{{\mathsf{Decomp}}}
\def\share{\mathsf{Share}}
\def\HomShare{\mathsf{HomShare}}
\def\rec{\mathsf{HomRec}}
\def\intFFT{\mathsf{IntFFT2D}}
\def\intIFFT{\mathsf{IntIFFT2D}}

\def\hatu{{\hat{u}}}
\def\hats{{\hat{s}}}
\def\hatw{{\hat{w}}}
\def\ovu{{\overline{u}}}
\def\ovw{{\overline{w}}}
\def\hatU{{\hat{U}}}
\def\hatR{{\hat{R}}}
\def\hatW{{\hat{W}}}
\def\hatbu{{\hat{\bu}}}
\def\hatbr{{\hat{\br}}}
\def\hatbw{{\hat{\bw}}}
\def\hatbv{{\hat{\bv}}}
\def\hatbx{{\hat{\bx}}}
\def\hatby{{\hat{\by}}}
\def\hatbs{{\hat{\bs}}}
\def\bpt{b}
\def\etarot{\eta_{\rm rot}}
\def\etamult{\eta_{\rm mult}}
\def\aaprox{\alpha}
\def\fe{{f_{\rm E}}}
\def\fd{{f_{\rm D}}}
\def\ee{{\varepsilon}}
\def\bee{{\bm{\varepsilon}}}
\def\bpi{{\bm{\pi}}}
\def\etal{et~al.\ }
\def\clgq{\lceil\lg{q}\rceil}
\def\clgp{\lceil\lg{\p}\rceil}
\def\clgpntt{\lceil\lg{\pntt}\rceil}
\def\clgpfft{\lceil\lg{\pfft}\rceil}
\def\clgpenc{\lceil\lg{\penc}\rceil}

\algnewcommand\algorithmicforeach{\textbf{for each}}
\algdef{S}[FOR]{ForEach}[1]{\algorithmicforeach\ #1\ \algorithmicdo}

\appendix
\section{Appendix: Correctness for Eq.~(17) to Eq.~(19)}
For some input $\hatbu$, weight vector $\hatbw$, secret share
$\hatbs_{B}$, HSS modulus $\pass$, NTT modulus $\pntt$, we have
\begin{align}
  &(\INTT(\hatbu\circ\hatbw-\hatbs_{B})\bmod{\pass})_{k}&\\
  &=\sum_{0}^{n_{f}-1}((\hatu_{i}\cdot \hatw_{i} - \hats_{B, i})\bmod{\pass})\omega^{ik}\bmod{\pntt}&\\
  &=\Big(\sum_{0}^{n_{f}-1}((\hatu_{i}\cdot \hatw_{i} - \hats_{B, i})\bmod{\pass})\omega^{ik}\bmod{\pntt}\Big)\bmod{\pass}&
\end{align}
Assuming $\pass\geq\pntt$, we know that for any $x$, $x\bmod{\pass}\bmod{\pntt}\equiv x\bmod{\pntt}$.
\begin{align}
  &=\Big(\sum_{0}^{n_{f}-1}((\hatu_{i}\cdot \hatw_{i})\bmod{\pass} - (\hats_{B, i})\bmod{\pass})\omega^{ik}\nonumber&\\
  &                                               \hspace{4.5cm}\bmod{\pntt}\Big)\bmod{\pass}&\\
  &=\Big(\sum_{0}^{n_{f}-1}((\hatu_{i}\cdot \hatw_{i})\bmod{\pntt} - (\hats_{B, i})\bmod{\pntt})\omega^{ik}\nonumber&\\
  &                                               \hspace{4.5cm}\bmod{\pntt}\Big)\bmod{\pass}&\\
  &=\Big( \sum_{0}^{n_{f}-1}(\hatu_{i}\cdot \hatw_{i}\bmod{\pntt})\omega^{ik}\bmod{\pntt}\Big)\bmod{\pntt} &\nonumber\\
  &\hspace{1.5em}-\Big(\sum_{0}^{n_{f}-1}(\hats_{B, i}\bmod{\pntt})\omega^{ik}\bmod{\pntt}\Big)\bmod{\pntt}&\\
  &=(\INTT(\hatbu\circ\hatbw)-\INTT(\hatbs_{B}))\bmod{\pass}&\\
  &=((\bu*\bw - \bs_{B})\bmod{\pntt})\bmod{\pass},&
\end{align}
To remove the additive secret sharing, observe that
\begin{align}
  &((\bu*\bw - \bs_{B})\bmod{\pntt}\bmod{\pass} + \bs_{B})\bmod{\pntt}&\\
  &= \bu*\bw\bmod{\pntt} - \bs_{B}\bmod{\pntt}+\bs_{B}\bmod{\pntt}&\\
  &= \bu*\bw\bmod{\pntt},&
\end{align}
and this addition can be computed homomorphically using any additive
homomorphic encryption scheme.


\section{Appendix: Neural Architectures}
\begin{figure}[!h]
  \centering
  \footnotesize
  \fbox{\parbox{\columnwidth}{
  \begin{enumerate}
    \item {\it{Convolution}}: input image $3\times 32\times 32$, weight matrix
      $64\times 3\times 3$, number of output channels
      64: \\$\zr^{64\times 32\times 32}\leftarrow \sum \zr^{3\times 32\times
      32}* \zr^{64\times 3\times 3}$ + ReLU.
    \item {\it{Convolution}}: input image $64\times 32\times 32$, weight matrix
      $64\times 3\times 3$, number of output channels
      64: \\$\zr^{64\times 32\times 32}\leftarrow \zr^{64\times 32\times
      32}* \zr^{64\times 3\times 3}$ + ReLU.
    \item {\it{Average Pooling}}: Outputs $\zr^{64\times 16\times 16}$.
    \item {\it{Convolution}}: input image $64\times 16\times 16$, weight matrix
      $64\times 3\times 3$, number of output channels
      64: \\$\zr^{64\times 16\times 16}\leftarrow \zr^{64\times 16\times
      16}* \zr^{64\times 3\times 3}$ +ReLU.
    \item {\it{Convolution}}: same as 6).
    \item {\it{Average Pooling}}: Outputs $\zr^{64\times 8\times 8}$
    \item {\it{Convolution}}: input image $64\times 8\times 8$, weight matrix
      $64\times 3\times 3$, number of output channels
      64: $\zr^{64\times 8\times 8}\leftarrow \zr^{64\times 8\times
      8}* \zr^{64\times 3\times 3}$ + ReLU.
    \item {\it{Convolution}}: input image $64\times 8\times 8$, weight matrix
      $64\times 1\times 1$, number of output channels
      64: $\zr^{64\times 8\times 8}\leftarrow \zr^{64\times 8\times
      8}* \zr^{64\times 1\times 1}$ + ReLU.
    \item {\it{Fully Connected}}: Outputs the classification result
      $\zr^{10\times 1}\leftarrow\zr^{10\times 1024}\cdot \zr^{1024\times 1}$
  \end{enumerate}
}}
  \caption{The neural architecture from Fig.~13 in [22]}
  \label{fig:fdcnn_architecture}
\end{figure}

\begin{figure}[!h]
  \centering
  \footnotesize
  \fbox{\parbox{\columnwidth}{
  \begin{enumerate}
    \item {\it{Convolution}}: input image $3\times 32\times 32$, weight matrix
      $32\times 3\times 3$, number of output channels
      32: \\$\zr^{32\times 32\times 32}\leftarrow \sum \zr^{3\times 32\times
      32}* \zr^{32\times 3\times 3}$ + ReLU.
    \item {\it{Convolution}}: input image $32\times 32\times 32$, weight matrix
      $64\times 3\times 3$, number of output channels
      64: \\$\zr^{64\times 32\times 32}\leftarrow \zr^{32\times 32\times
      32}* \zr^{64\times 3\times 3}$ + Square Activation.
    \item {\it{Average Pooling}}: Outputs $\zr^{64\times 16\times 16}$.
    \item {\it{Convolution}}: input image $64\times 16\times 16$, weight matrix
      $64\times 3\times 3$, number of output channels
      64: \\$\zr^{64\times 16\times 16}\leftarrow \zr^{64\times 16\times
      16}* \zr^{64\times 3\times 3}$ + ReLU.
    \item {\it{Convolution}}: input image $64\times 16\times 16$, weight matrix
      $64\times 3\times 3$, number of output channels
      64: \\$\zr^{64\times 16\times 16}\leftarrow \zr^{64\times 16\times
      16}* \zr^{64\times 3\times 3}$ + Square Activation.
    \item {\it{Average Pooling}}: Outputs $\zr^{64\times 8\times 8}$
    \item {\it{Convolution}}: input image $64\times 8\times 8$, weight matrix
      $64\times 3\times 3$, number of output channels
      64: $\zr^{64\times 8\times 8}\leftarrow \zr^{64\times 8\times
      8}* \zr^{64\times 3\times 3}$ + ReLU.
    \item {\it{Convolution}}: input image $64\times 8\times 8$, weight matrix
      $64\times 1\times 1$, number of output channels
      64: $\zr^{64\times 8\times 8}\leftarrow \zr^{64\times 8\times
      8}* \zr^{64\times 1\times 1}$ + ReLU.
    \item {\it{Fully Connected}}: Outputs the classification result
      $\zr^{10\times 1}\leftarrow\zr^{10\times 1024}\cdot \zr^{1024\times 1}$
  \end{enumerate}
}}
  \caption{The modified neural architecture from Fig.~13 in [22], where the 
  number of channels in the first convolution layer is reduced  by half, and the second
  and fifth activation function is replaced with square activation.}
  \label{fig:fdcnn_architecture}
\end{figure}

\begin{figure}[!h]
  \centering
  \footnotesize
  \fbox{\parbox{\columnwidth}{
  \begin{enumerate}
    \item {\it{Convolution}}: input image $3\times 32\times 32$, weight matrix
      $16\times 3\times 3$, number of output channels
      16: \\$\zr^{16\times 32\times 32}\leftarrow \sum \zr^{3\times 32\times
      32}* \zr^{16\times 3\times 3}$ + Batch Normalization + Binary Activation.
    \item {\it{Convolution}}: input image $16\times 32\times 32$, weight matrix
      $16\times 3\times 3$, number of output channels
      16: \\$\zr^{16\times 32\times 32}\leftarrow \sum \zr^{16\times 32\times
      32}* \zr^{16\times 3\times 3}$ + Batch Normalization + Binary Activation.     
    \item Same as Layer 2.
    \item {\it{Average Pooling}}: Outputs $\zr^{16\times 16\times 16}$.
    \item {\it{Convolution}}: input image $16\times 16\times 16$, weight matrix
      $32\times 3\times 3$, number of output channels
      32: \\$\zr^{32\times 16\times 16}\leftarrow \sum \zr^{16\times 16\times
      16}* \zr^{32\times 3\times 3}$ + Batch Normalization + Binary Activation.     
    \item {\it{Convolution}}: input image $32\times 16\times 16$, weight matrix
      $32\times 3\times 3$, number of output channels
      32: \\$\zr^{32\times 16\times 16}\leftarrow \sum \zr^{16\times 16\times
      16}* \zr^{32\times 3\times 3}$ + Batch Normalization + Binary Activation.     
    \item Same as Layer 6.
    \item {\it{Average Pooling}}: Outputs $\zr^{32\times 8\times 8}$
    \item {\it{Convolution}}: input image $32\times 8\times 8$, weight matrix
      $48\times 3\times 3$, number of output channels
      48: \\$\zr^{48\times 6\times 6}\leftarrow \sum \zr^{32\times 8\times
      8}* \zr^{48\times 3\times 3}$ + Batch Normalization + Binary Activation.     
    \item {\it{Convolution}}: input image $48\times 6\times 6$, weight matrix
      $48\times 3\times 3$, number of output channels
      48: \\$\zr^{48\times 4\times 4}\leftarrow \sum \zr^{48\times 6\times
      6}* \zr^{48\times 3\times 3}$ + Batch Normalization + Binary Activation.     
    \item {\it{Convolution}}: input image $48\times 4\times 4$, weight matrix
      $64\times 3\times 3$, number of output channels
      64: \\$\zr^{64\times 2\times 2}\leftarrow \sum \zr^{48\times 4\times
      4}* \zr^{64\times 3\times 3}$ + Batch Normalization + Binary Activation.     
    \item {\it{Average Pooling}}: Outputs $\zr^{64\times 1\times 1}$
    \item {\it{Fully Connected}}: Outputs the classification result
      $\zr^{10\times 1}\leftarrow\zr^{10\times 64}\cdot \zr^{64\times 1}$
  \end{enumerate}
}}
  \caption{BC2 from [26]}
  \label{fig:fdcnn_architecture}
\end{figure}

\begin{figure}[!h]
  \centering
  \footnotesize
  \fbox{\parbox{\columnwidth}{
  \begin{enumerate}
    \item {\it{Convolution}}: input image $3\times 32\times 32$, weight matrix
      $16\times 3\times 3$, number of output channels
      16: \\$\zr^{16\times 32\times 32}\leftarrow \sum \zr^{3\times 32\times
      32}* \zr^{16\times 3\times 3}$ + Batch Normalization + Binary Activation.
    \item {\it{Convolution}}: input image $16\times 32\times 32$, weight matrix
      $32\times 3\times 3$, number of output channels
      32: \\$\zr^{32\times 32\times 32}\leftarrow \sum \zr^{16\times 32\times
      32}* \zr^{32\times 3\times 3}$ + Batch Normalization + Binary Activation.     
    \item {\it{Convolution}}: input image $32\times 32\times 32$, weight matrix
      $32\times 3\times 3$, number of output channels
      32: \\$\zr^{32\times 32\times 32}\leftarrow \sum \zr^{32\times 32\times
      32}* \zr^{32\times 3\times 3}$ + Batch Normalization + Binary Activation.     
    \item Same as Layer 3.
    \item {\it{Average Pooling}}: Outputs $\zr^{32\times 16\times 16}$.
    \item {\it{Convolution}}: input image $32\times 16\times 16$, weight matrix
      $48\times 3\times 3$, number of output channels
      48: \\$\zr^{32\times 16\times 16}\leftarrow \sum \zr^{48\times 16\times
      16}* \zr^{32\times 3\times 3}$ + Batch Normalization + Binary Activation.     
    \item {\it{Convolution}}: input image $48\times 16\times 16$, weight matrix
      $64\times 3\times 3$, number of output channels
      64: \\$\zr^{64\times 16\times 16}\leftarrow \sum \zr^{48\times 16\times
      16}* \zr^{64\times 3\times 3}$ + Batch Normalization + Binary Activation.     
    \item {\it{Convolution}}: input image $64\times 16\times 16$, weight matrix
      $80\times 3\times 3$, number of output channels
      80: \\$\zr^{80\times 16\times 16}\leftarrow \sum \zr^{64\times 16\times
      16}* \zr^{80\times 3\times 3}$ + Batch Normalization + Binary Activation.     
    \item {\it{Average Pooling}}: Outputs $\zr^{80\times 8\times 8}$
    \item {\it{Convolution}}: input image $80\times 8\times 8$, weight matrix
      $96\times 3\times 3$, number of output channels
      96: \\$\zr^{96\times 6\times 6}\leftarrow \sum \zr^{80\times 8\times
      8}* \zr^{96\times 3\times 3}$ + Batch Normalization + Binary Activation.     
    \item {\it{Convolution}}: input image $96\times 6\times 6$, weight matrix
      $96\times 3\times 3$, number of output channels
      96: \\$\zr^{96\times 4\times 4}\leftarrow \sum \zr^{96\times 6\times
      6}* \zr^{96\times 3\times 3}$ + Batch Normalization + Binary Activation.     
    \item {\it{Convolution}}: input image $96\times 4\times 4$, weight matrix
      $128\times 3\times 3$, number of output channels
      128: \\$\zr^{128\times 2\times 2}\leftarrow \sum \zr^{96\times 4\times
      4}* \zr^{128\times 3\times 3}$ + Batch Normalization + Binary Activation.     
    \item {\it{Average Pooling}}: Outputs $\zr^{128\times 1\times 1}$
    \item {\it{Fully Connected}}: Outputs the classification result
      $\zr^{10\times 1}\leftarrow\zr^{10\times 128}\cdot \zr^{128\times 1}$
  \end{enumerate}
}}
  \caption{BC3 from [26]}
  \label{fig:fdcnn_architecture}
\end{figure}

\begin{figure}[!h]
  \centering
  \footnotesize
  \fbox{\parbox{\columnwidth}{
  \begin{enumerate}
    \item {\it{Convolution}}: input image $3\times 32\times 32$, weight matrix
      $32\times 3\times 3$, number of output channels
      32: \\$\zr^{32\times 32\times 32}\leftarrow \sum \zr^{3\times 32\times
      32}* \zr^{32\times 3\times 3}$ + Batch Normalization + Binary Activation.
    \item {\it{Convolution}}: input image $32\times 32\times 32$, weight matrix
      $32\times 3\times 3$, number of output channels
      32: \\$\zr^{32\times 32\times 32}\leftarrow \sum \zr^{32\times 32\times
      32}* \zr^{32\times 3\times 3}$ + Batch Normalization + Binary Activation.     
    \item {\it{Convolution}}: input image $32\times 32\times 32$, weight matrix
      $48\times 3\times 3$, number of output channels
      48: \\$\zr^{48\times 32\times 32}\leftarrow \sum \zr^{32\times 32\times
      32}* \zr^{48\times 3\times 3}$ + Batch Normalization + Binary Activation.     
    \item {\it{Convolution}}: input image $48\times 32\times 32$, weight matrix
      $64\times 3\times 3$, number of output channels
      64: \\$\zr^{64\times 32\times 32}\leftarrow \sum \zr^{48\times 32\times
      32}* \zr^{64\times 3\times 3}$ + Batch Normalization + Binary Activation.     
    \item {\it{Convolution}}: input image $64\times 32\times 32$, weight matrix
      $64\times 3\times 3$, number of output channels
      64: \\$\zr^{64\times 32\times 32}\leftarrow \sum \zr^{64\times 32\times
      32}* \zr^{64\times 3\times 3}$ + Batch Normalization + Binary Activation.     
    \item {\it{Average Pooling}}: Outputs $\zr^{64\times 16\times 16}$.
    \item {\it{Convolution}}: input image $64\times 16\times 16$, weight matrix
      $80\times 3\times 3$, number of output channels
      80: \\$\zr^{80\times 16\times 16}\leftarrow \sum \zr^{64\times 16\times
      16}* \zr^{80\times 3\times 3}$ + Batch Normalization + Binary Activation.     
    \item {\it{Convolution}}: input image $80\times 16\times 16$, weight matrix
      $80\times 3\times 3$, number of output channels
      80: \\$\zr^{80\times 16\times 16}\leftarrow \sum \zr^{80\times 16\times
      16}* \zr^{80\times 3\times 3}$ + Batch Normalization + Binary Activation.     
    \item Same as Layer 8.
    \item Same as Layer 8.
    \item {\it{Average Pooling}}: Outputs $\zr^{80\times 8\times 8}$
    \item {\it{Convolution}}: input image $80\times 8\times 8$, weight matrix
      $128\times 3\times 3$, number of output channels
      128: \\$\zr^{128\times 6\times 6}\leftarrow \sum \zr^{80\times 8\times
      8}* \zr^{128\times 3\times 3}$ + Batch Normalization + Binary Activation.     
    \item {\it{Convolution}}: input image $128\times 6\times 6$, weight matrix
      $128\times 3\times 3$, number of output channels
      128: \\$\zr^{128\times 4\times 4}\leftarrow \sum \zr^{128\times 6\times
      6}* \zr^{128\times 3\times 3}$ + Batch Normalization + Binary Activation.     
    \item {\it{Convolution}}: input image $128\times 4\times 4$, weight matrix
      $128\times 3\times 3$, number of output channels
      128: \\$\zr^{128\times 2\times 2}\leftarrow \sum \zr^{128\times 4\times
      4}* \zr^{128\times 3\times 3}$ + Batch Normalization + Binary Activation.     
    \item {\it{Average Pooling}}: Outputs $\zr^{128\times 1\times 1}$
    \item {\it{Fully Connected}}: Outputs the classification result
      $\zr^{10\times 1}\leftarrow\zr^{10\times 128}\cdot \zr^{128\times 1}$
  \end{enumerate}
}}
  \caption{BC4 from [26]}
  \label{fig:fdcnn_architecture}
\end{figure}

\begin{figure}[!h]
  \centering
  \footnotesize
  \fbox{\parbox{\columnwidth}{
  \begin{enumerate}
    \item {\it{Convolution}}: input image $3\times 32\times 32$, weight matrix
      $32\times 3\times 3$, number of output channels
      32: \\$\zr^{32\times 32\times 32}\leftarrow \sum \zr^{3\times 32\times
      32}* \zr^{32\times 3\times 3}$ + Batch Normalization + Binary Activation.
    \item {\it{Convolution}}: input image $32\times 32\times 32$, weight matrix
      $32\times 3\times 3$, number of output channels
      32: \\$\zr^{32\times 32\times 32}\leftarrow \sum \zr^{32\times 32\times
      32}* \zr^{32\times 3\times 3}$ + Batch Normalization + Binary Activation.     
    \item Same as Layer 2.
    \item Same as Layer 2.
    \item {\it{Convolution}}: input image $32\times 32\times 32$, weight matrix
      $48\times 3\times 3$, number of output channels
      48: \\$\zr^{48\times 32\times 32}\leftarrow \sum \zr^{32\times 32\times
      32}* \zr^{48\times 3\times 3}$ + Batch Normalization + Binary Activation.     
    \item {\it{Convolution}}: input image $48\times 32\times 32$, weight matrix
      $48\times 3\times 3$, number of output channels
      48: \\$\zr^{48\times 32\times 32}\leftarrow \sum \zr^{48\times 32\times
      32}* \zr^{48\times 3\times 3}$ + Batch Normalization + Binary Activation.     
    \item {\it{Average Pooling}}: Outputs $\zr^{48\times 16\times 16}$.
    \item {\it{Convolution}}: input image $48\times 16\times 16$, weight matrix
      $80\times 3\times 3$, number of output channels
      80: \\$\zr^{80\times 16\times 16}\leftarrow \sum \zr^{48\times 16\times
      16}* \zr^{80\times 3\times 3}$ + Batch Normalization + Binary Activation.     
    \item {\it{Convolution}}: input image $80\times 16\times 16$, weight matrix
      $80\times 3\times 3$, number of output channels
      80: \\$\zr^{80\times 16\times 16}\leftarrow \sum \zr^{80\times 16\times
      16}* \zr^{80\times 3\times 3}$ + Batch Normalization + Binary Activation.     
    \item Same as Layer 7.
    \item Same as Layer 7.
    \item Same as Layer 7.
    \item Same as Layer 7.
    \item {\it{Average Pooling}}: Outputs $\zr^{80\times 8\times 8}$
    \item {\it{Convolution}}: input image $80\times 8\times 8$, weight matrix
      $128\times 3\times 3$, number of output channels
      128: \\$\zr^{128\times 6\times 6}\leftarrow \sum \zr^{80\times 8\times
      8}* \zr^{128\times 3\times 3}$ + Batch Normalization + Binary Activation.     
    \item {\it{Convolution}}: input image $128\times 8\times 8$, weight matrix
      $128\times 3\times 3$, number of output channels
      128: \\$\zr^{128\times 8\times 8}\leftarrow \sum \zr^{128\times 8\times
      8}* \zr^{128\times 3\times 3}$ + Batch Normalization + Binary Activation.     
    \item Same as 16.
    \item {\it{Convolution}}: input image $128\times 8\times 8$, weight matrix
      $128\times 3\times 3$, number of output channels
      128: \\$\zr^{128\times 6\times 6}\leftarrow \sum \zr^{128\times 8\times
      8}* \zr^{128\times 3\times 3}$ + Batch Normalization + Binary Activation.     
    \item {\it{Convolution}}: input image $128\times 6\times 6$, weight matrix
      $128\times 3\times 3$, number of output channels
      128: \\$\zr^{128\times 4\times 4}\leftarrow \sum \zr^{128\times 6\times
      6}* \zr^{128\times 3\times 3}$ + Batch Normalization + Binary Activation.     
    \item {\it{Convolution}}: input image $128\times 4\times 4$, weight matrix
      $128\times 3\times 3$, number of output channels
      128: \\$\zr^{128\times 2\times 2}\leftarrow \sum \zr^{128\times 4\times
      4}* \zr^{128\times 3\times 3}$ + Batch Normalization + Binary Activation.     
    \item {\it{Average Pooling}}: Outputs $\zr^{128\times 1\times 1}$
    \item {\it{Fully Connected}}: Outputs the classification result
      $\zr^{10\times 1}\leftarrow\zr^{10\times 128}\cdot \zr^{128\times 1}$
  \end{enumerate}
}}
  \caption{BC5 from [26]}
  \label{fig:fdcnn_architecture}
\end{figure}